\documentclass[%
 aip,
 jmp,%
 amsmath,amssymb,
 reprint,%
groupedaddress,
]{revtex4-1}

\usepackage{graphicx}
\usepackage{dcolumn}
\usepackage{bm}
\usepackage{hyperref} 
\usepackage{color}
\usepackage{multirow,makecell,booktabs}
\usepackage{enumitem}

\begin{document}


\title{Improved lifetime of a high spin polarization superlattice photocathode
}

\author{Jai Kwan Bae}
\author{Alice Galdi}
\author{Luca Cultrera}
\author{Frank Ikponmwen}


\author{Jared Maxson}
\author{Ivan Bazarov}

\affiliation{Cornell Laboratory for Accelerator-Based Sciences and Education, Cornell University, Ithaca, NY 14853, USA}

\date{\today}

\begin{abstract}
Negative Electron Affinity (NEA) activated surfaces are required to extract highly spin polarized electron beams from GaAs-based photocathodes, but they suffer extreme sensitivity to poor vacuum conditions that results in rapid degradation of quantum efficiency. We report on series of unconventional NEA activations on surfaces of bulk GaAs with Cs, Sb, and O$_2$ using different methods of oxygen exposure for optimizing photocathode performance. One order of magnitude improvement in lifetime with respect to the standard Cs-O$_2$ activation is achieved without significant loss on electron spin polarization and quantum efficiency by codepositing Cs, Sb, and O$_2$. A strained GaAs/GaAsP superlattice sample activated with the codeposition method demonstrated similar enhancement in lifetime near the photoemission threshold while maintaining 90\% spin polarization.

\end{abstract}

\maketitle

\section{Introduction}
Highly spin polarized electron beams with high currents are required by future nuclear physics facilities such as the Electron Ion Collider and the International Linear Collider.\cite{eic,lrpns,nsac} Photocathode sources capable of providing spin polarized electron beams at high currents ($\sim$ 50 mA)\cite{NASreport} for extended period of time which are robust against less than ideal vacuum conditions are required in order to reduce the cost and complexity of operating these new facilities. Long lifetime polarized electron sources are also of interest for electron microscopy technologies that exploits spin polarization to probe magnetization in materials and nanostructures at the nanoscale level. Additionally, bright spin polarized sources driven with a short pulse laser can enable time studies of magnetization dynamics.\cite{tr-spleem, srtem,Vollmer2003}
GaAs-based photocathodes are the present state of the art for the spin polarized electron sources because they can provide highly spin polarized electron beams with high efficiency.\cite{dbr} However, their short operational lifetime poses limits particularly for applications requiring high average current.\cite{Cardman2018}

GaAs was first shown to be an attractive electron photoemitter five decades ago via the discovery of negative electron affinity (NEA) activation on \emph{p}-type samples.\cite{Scheer1965}
When the GaAs surface is exposed to cesium vapor (electropositive metal atoms), a strong dipole layer is formed on the surface that lowers the electron affinity, which is defined as the energy difference between the vacuum level and conduction band minimum.\cite{Liu2017} If the GaAs is $p$-doped, the effective electron affinity can become negative because of the downward band bending near the surface. Thus, in NEA conditions, electrons that have relaxed down to the bottom of the conduction band after photoexcitation can still escape into vacuum when they reach the vacuum interface. The combination of NEA and other characteristics of GaAs-based photocathodes, such as high electron mobility, results in a high quantum efficiency (QE) at the band gap photon energy of about 1.4 eV.\cite{Pierce1976} Later, it was demonstrated that the introduction of an oxidant, such as oxygen or NF$_3$, during cesiation can yield an even stronger dipole moment layer and achieve larger NEA and hence larger QE.\cite{Su1983,Ciccacci1991}

Spin polarized electron photoemission from GaAs is achieved by exploiting the quantum mechanical selection rule that conserves the total angular momentum during photoexcitation with circularly polarized light.\cite{Pierce1976} Since electrons need to be excited only from the top of the valence band for a high spin polarization, NEA is required.
Due to the degenerate light-hole and heavy-hole band states in the P$_{3/2}$ valence band, the theoretical limit of spin polarization of NEA bulk GaAs is 50\%.
Additionally, various spin relaxation mechanisms limit spin polarization from bulk GaAs to $\sim$ 35 \% at room temperature.\cite{Liu2017} A number of approaches have been proposed to overcome this limit,\cite{Clayburn2013,McCarter2014} where exerting lattice strain on GaAs aimed at breaking the valence band degeneracy has been the most successful. This method exhibits a high spin polarization of 90\%, but suffers from significantly decreased QE of 0.07\%.\cite{strained} In an extension of this approach, GaAs/GaAsP multi-layer superlattices structures with a distributed Bragg reflector were shown to improve the QE up to 6.4\% with a spin polarization of 84\%.\cite{dbr}

NEA surface of GaAs suffers from extreme vacuum sensitivity because traditional activation layer materials such as Cs-O$_2$ and Cs-NF$_3$ form about a monolayer that is weakly bound to the surface and is chemically reactive.\cite{Kuriki2011,Chanlek2014}
 Enhancement of activation layer chemical stability has been shown by using two alkali species, Cs and Li, during activation.\cite{Mulhollan2008,Sun2009,kurichiyanil2019test} 
While the chemical reactivity is responsible for QE degradation when no photoemission is occurring (dark lifetime), the ion back-bombardment effect also limits the lifetime of the GaAs-based photocathodes during beam extraction.
Deflecting the electron beam near the emission site was proposed to counteract this mechanism and showed an improvement in lifetime of both GaAs and alkali antimonide photocathodes.
\cite{Grames2011,Mammei2013,Rahman2019,Cultrera2011} 

Although $n$-type Cs$_3$Sb and CsI were demonstrated to be alternative activation materials at the time of the discovery of NEA activation,\cite{Sonnenberg1969, Sonnenberg1969Cs,Hagino1969,Zhao1993,Guo1996} they have generally been avoided in the photocathode community because the semiconductor activation layer is thicker, and can potentially decrease spin polarization.
Recent studies reported successful NEA activation on bulk GaAs using Cs$_2$Te,\cite{Sugiyama2011a,Uchida,kuriki} a semiconductor known for being a robust solar blind photocathode material itself.\cite{robust} It was demonstrated that Cs$_2$Te activation layer can improve the charge extraction lifetime by a factor of 5 without negatively affecting the spin polarization.\cite{gaas_cste} Yet, all experiments were done with bulk GaAs which only has up to 40\% spin polarization at the room temperature, and semiconductor activation of high spin polarization photocathodes (like GaAs/GaAsP superlattice) and its effect on the degree of polarization have not been demonstrated.



\begin{figure}
	\includegraphics*[width=230pt]{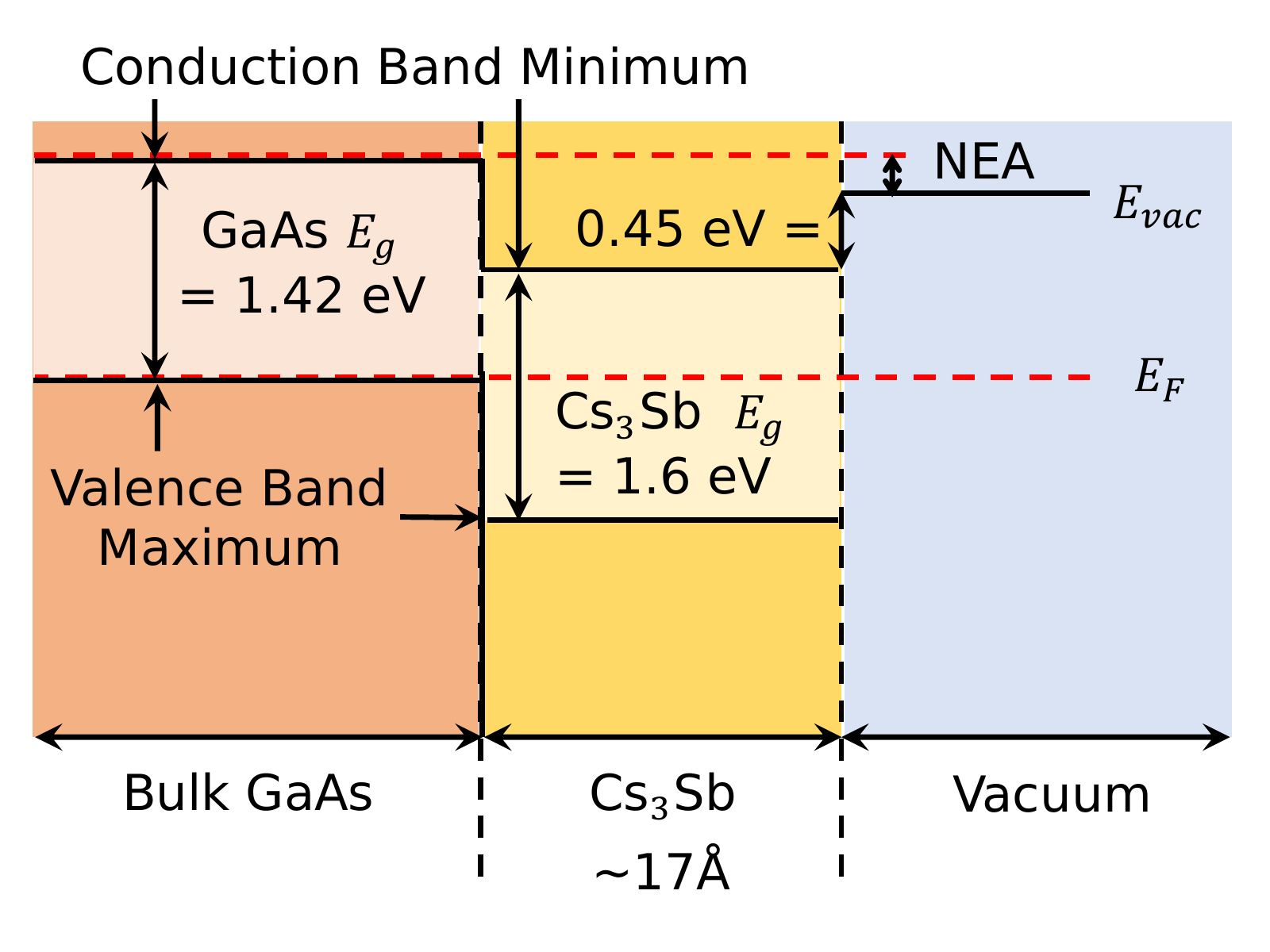}
	\caption{Energy band diagram of GaAs activated with Cs$_3$Sb coating. An alternative activation layer needs to satisfy two conditions to achieve NEA. (i) The energy difference between the vacuum level $E_{vac}$ and Fermi level $E_F$ should be smaller than the GaAs band gap. (ii) The band gap of activation layer should be larger than the GaAs band gap so that the activation layer is transparent for photon energy near the GaAs band gap.}
	\label{fig_band}
\end{figure}

Based on the heterojunction model, a semiconductor layer capable of NEA activation on $p$-type GaAs should satisfy two conditions (see Fig.~\ref{fig_band}): (i) the energy difference between the Fermi level and vacuum level should be smaller than the band gap of GaAs (1.42 eV) to achieve NEA, and (ii) the band gap of activation layer should be greater than that of GaAs to ensure transparency to photons with the GaAs band gap energy.
Cs$_3$Sb has a band gap of 1.6 eV and small electron affinity of 0.45 eV.
Such a small electron affinity hints at the possibility of NEA activation on GaAs using Cs$_3$Sb: as shown in Fig.~\ref{fig_band}, Cs$_3$Sb satisfies the two conditions mentioned above without any doping control. 

Recently, Ref.~\onlinecite{LucaPaper} reported NEA activation on GaAs surfaces with Cs, Sb, and O$_2$ codeposited layer and changes in photocathode parameters as the thickness was varied.
In this work, we demonstrate NEA conditions achieved on bulk $p$-type GaAs with Cs-Sb using various methods of oxygen exposure and its effect on photocathode parameters such as the lifetime, QE, and the degree of spin polarization.
Then, a GaAs/GaAsP superlattice sample was activated with one of the methods (codeposition of Cs, Sb, and O$_2$) and compared with typical Cs-O$_2$ activation. We find a significant improvement in the lifetime at 780 nm while preserving 90\% spin polarization.


\section{Growth}

$10\times10$ mm$^2$ samples are prepared by cutting highly \emph{p}-doped (Zn $5\times 10^{18} \text{ cm}^{-3}$) GaAs (100) wafers in air with a diamond scriber. Samples are then solvent cleaned with isopropanol and rinsed in de-ionized water. Wet-etching was later performed for each sample with 1\% HF for 30 s to remove the native oxide layer with minimal surface damage.\cite{ReiVilar2005,Feng2019_etching} Samples were finally rinsed again in de-ionized water and then moved into vacuum. 
The growth chamber has Cs and Sb effusion cells installed under ultra-high vacuum (UHV) of $\sim 10^{-9}$ Torr. Shutters in front of each effusion cell control the flux on the sample surface. A leak valve connected to an oxygen bottle was used to leak oxygen into the chamber during the growths. Total pressure is monitored with a cold cathode gauge. GaAs samples were heat cleaned at $\sim 500$ $^\circ$C for about 12 hours. Then temperature was lowered to about 130 $^\circ$C for the growth.
The same heat cleaning procedure was employed to prepare the reference sample which was later activated with Cs and O$_2$ at room temperature.

\begin{figure}
	\includegraphics*[width=230pt]{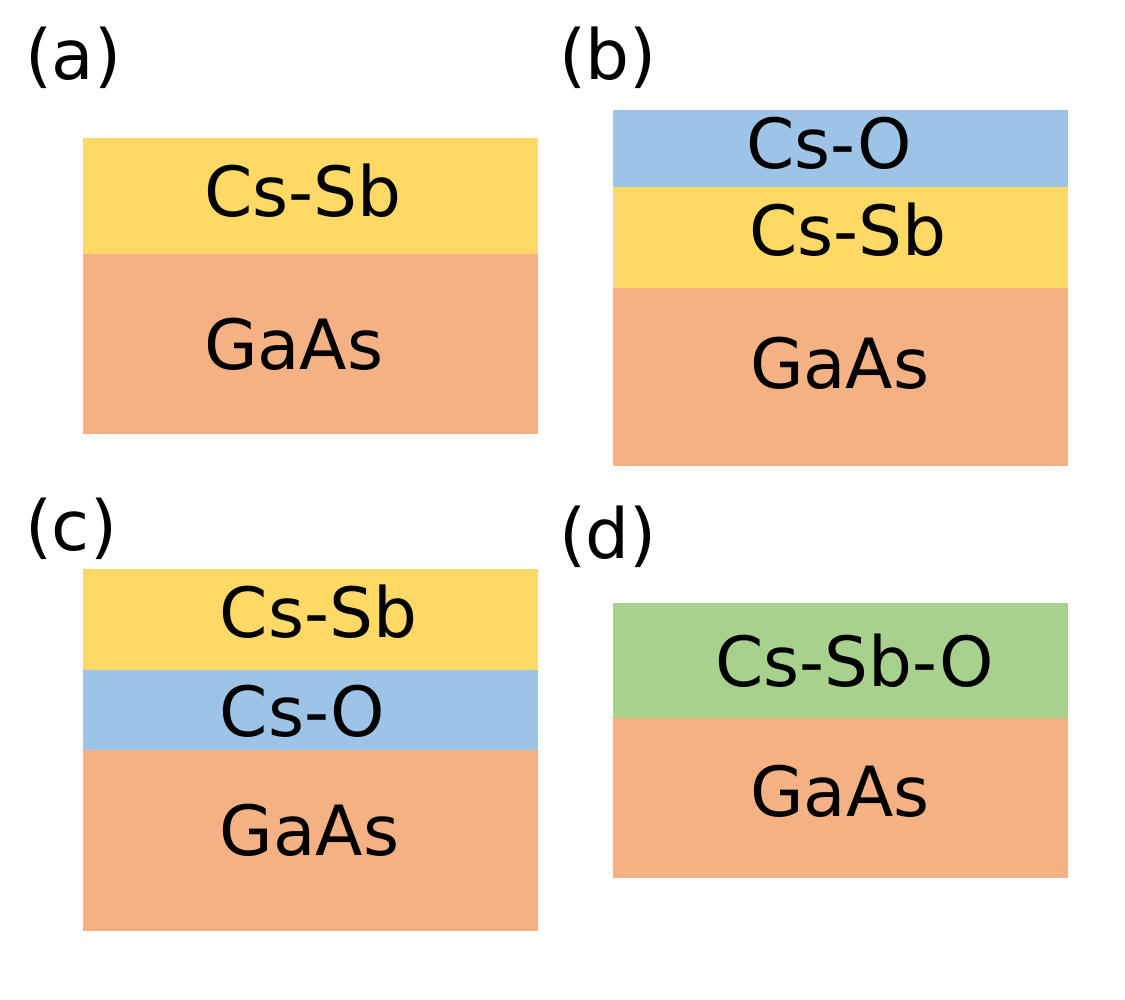}
	\caption{Deposition order of different oxygen exposure methods used in Cs-Sb semiconductor activation. (a)-(d) represents methods (a)-(d) in the text.}
	\label{fig_growth_schematics}
\end{figure}

\begin{figure}
	\includegraphics*[width=230pt]{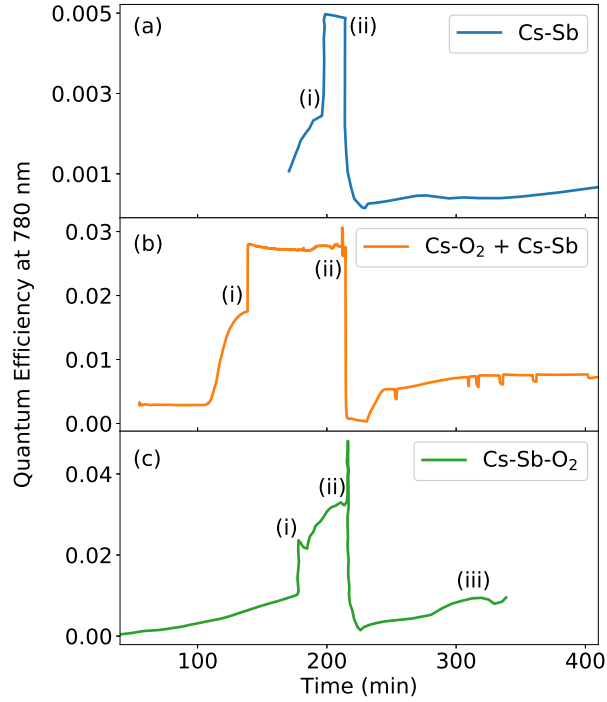}
	\caption{Quantum Efficiency of GaAs photocathodes during the thin film growths. The initial QE at 780 nm is not zero due to residual cesium vapor in the growth chamber. The Cs shutter and Sb shutter are opened at (i) and (ii), respectively. (a) Oxygen was not used during the growth. (b) Oxygen was leaked before deposition of Sb. (c) Cs, Sb, and O$_2$ are codeposited during the growths. Oxygen leak valve was closed at (iii).}
	\label{fig_growth}
\end{figure}

A 780 nm diode laser was used to excite the samples\cite{Feng2019} that were biased at -18 V during the growth. This wavelength is at the NEA GaAs threshold. The drain current was measured via a lock-in amplifier locked to the frequency of an optical chopper used to modulate the laser light.
Samples were grown with five different methods:
\begin{enumerate}[label=(\alph*)]
	\item Cs and Sb are codeposited on GaAs without oxygen exposure.
	\item Cs-O$_2$ codeposition was additionally performed after Cs-Sb codeposition to study the effect of Cs-O$_2$ dipole layer on the surface.
	\item Cs and O$_2$ are codeposited before the growth Cs-Sb to study the effect of Cs-O$_2$ dipole layer at the interface between GaAs and Cs-Sb layer.
	\item Cs and Sb are codeposited under O$_2$ exposure throughout the growth procedure.
	\item Conventional codeposition of Cs and O$_2$ was performed for a control sample.
\end{enumerate}
Figure \ref{fig_growth_schematics} illustrates the layer order for samples grown with Sb.

In Fig.~\ref{fig_growth}, the QEs estimated during the different growth procedures of Cs-Sb on bulk GaAs with and without  the oxygen exposure are plotted.  
At the beginning of the growth procedures, we found the samples already photoemitting at 780 nm. We attribute this to the Cs vapors already present in the UHV chamber from previous growth experiments. As the Cs furnace temperature is raised with the shutter closed, the QE increases further, likely because Cs vapors can make it through the gap between the shutter and furnace. In Fig.~\ref{fig_growth}, (i) indicates the opening of the Cs shutter and (ii) is when the Sb shutter is opened. For all methods, the Sb shutter was left opened for 1000 seconds to deposit 2.5 \AA\ with a flux of $8.3 \times 10^{11}$ atoms/cm$^2$/s. Assuming Cs$_3$Sb single crystal structure, this amount of Sb corresponds to the total film thickness of 17\AA.\cite{Hagino1966} The Cs shutter was left open during the Sb layer deposition and the final cooling of the sample down to 50$^\circ$C. The photocurrent showed a sudden increase as the Cs shutter is opened and rapid decrease when Sb shutter is opened. 
For the samples of Fig. \ref{fig_growth} (a), no oxygen was leaked into the growth chamber (methods (a) and (b)). Oxygen was supplied with a partial pressure of $\sim 5 \times 10^{-9}$ Torr  in Fig.~\ref{fig_growth} (b) and (c); the leak valve was closed just before opening the Sb shutter in (b) (method (c)) and during cooling down (iii) in (c) (method (d)). %
The sample activated with method (b) was moved under UHV into another chamber (with a base pressure of $\sim 5 \times 10^{-11}$ Torr) to be further activated by exposing simultaneously to Cs and O$_2$ at room temperature. For the superlattice sample, Cs, Sb, and O$_2$ are codeposited for activation as in Fig.~\ref{fig_growth} (c).
Half the amount of Sb was deposited compared to the bulk samples (1.25 \AA\ Sb and 8.5 \AA\ total thickness) to maintain the spin polarization.\cite{LucaPaper}

\section{Results}


\begin{figure}
	\includegraphics*[width=230pt]{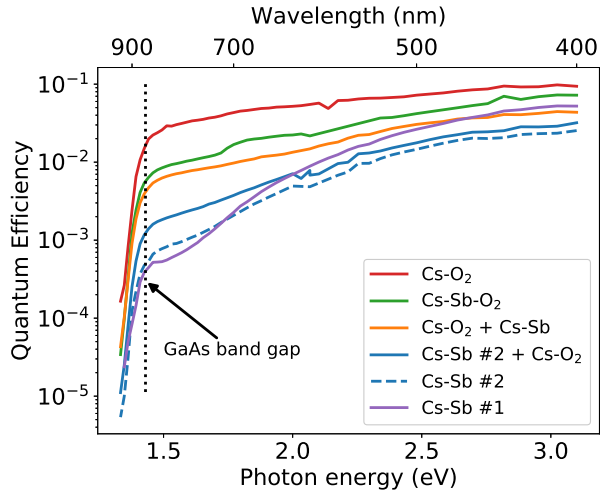}
	\caption{Spectral response of GaAs samples activated by Cs, Sb, and O$_2$ using different methods. All samples indicate NEA activation on the surface by photoemission at GaAs band gap energy (1.42 eV).
	}
	\label{fig_sr}
\end{figure}

The spectral response of samples with different growth methods are reported in Fig.~\ref{fig_sr}. 
Photoemission at the GaAs band gap photon energy (1.42 eV) confirms NEA achievement on the surface for all of the investigated activation methods. The reference sample, activated with Cs and O$_2$, has the highest QE across the measured spectral range. The sample activated with only Cs-Sb has the lowest QE in the infrared region (about 2 orders of magnitude lower than the Cs-O$_2$ activated sample at 780 nm). Exposing to Cs-O$_2$ after or before the growth of the Cs-Sb layer (methods (b) and (c)) increases the QE near the GaAs bandgap energy: in particular for the sample prepared with method (c), we observed an increase of QE by about an order of magnitude with respect to the bare Cs-Sb activated sample. Finally, the Cs-Sb-O$_2$ codeposited sample (method (d)) showed the highest QE near the photoemission threshold among Sb-containing samples. These results illustrate the significance of the Cs-O$_2$ dipole layer in enhancing the NEA at the GaAs interface.


	
\begin{figure}
	\includegraphics*[width=230pt]{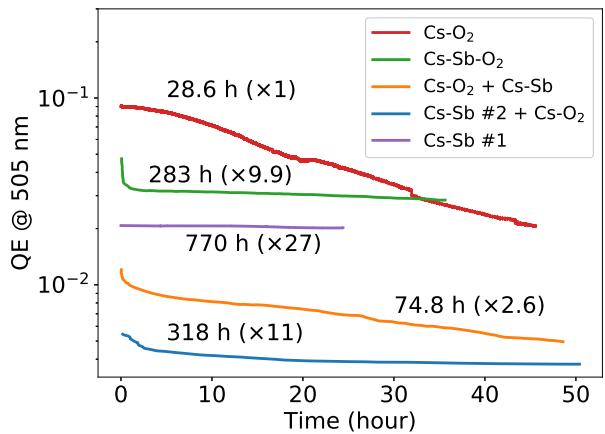}
	\caption{Quantum Efficiency degradation as a function of time. 505 nm laser was used to illuminate the sample. The number next to each curve is the lifetime calculated by fitting each curve to an exponential function. Improvement factors from the standard Cs-O$_2$ activation are in the parentheses.}
	\label{fig_lifetime}
\end{figure}

The robustness of the activating layers was compared by measuring the QE degradation over time. The lifetime of a photocathode is defined as the time QE takes to drop by a factor of $e$, the base of the natural logarithm.\cite{gaas_cste,Grames2011} 
Photocurrent was measured continuously in the range of 1 - 100 nA with a 505 nm diode laser ($\sim 20 \: \mu$W) in Fig.~\ref{fig_lifetime}. The Lifetime extracted from these measurements represent a convolution of the QE degradation from chemical poisoning (known as dark lifetime) and ion back-bombardment (known as charge lifetime).
A detailed study of dark lifetime and charge extraction lifetime of codeposited samples is reported in Ref.~\onlinecite{LucaPaper}.

From the data reported in Fig.~\ref{fig_lifetime}, it can be seen that Sb deposited samples have at least a factor of 2.6 improvement in lifetime compared to the sample activated with Cs and O$_2$. The longest lifetime (770 h) was observed when bare Cs-Sb is used for activation, but this method yielded the lowest efficiency near the threshold (Fig.~\ref{fig_sr}).
One order of magnitude increase in lifetime ($\sim$ 300 h) was observed when the samples are exposed to oxygen either after or during Cs-Sb growths.  
Table \ref{table2} reports estimated lifetimes at 780 nm.
Spectral response at 780 nm was measured before and after the QE degradation measurement in Fig.~\ref{fig_lifetime}, and the lifetime is calculated assuming an exponential decrease as a function of time between the two measured points. The Cs-Sb-O$_2$ activation showed the greatest improvement from Cs-O$_2$ activation by a factor of 6.8.
As opposed to Cs$_2$Te activation, which showed a rapid increase of the work function as a function of time,\cite{Sugiyama2011a} the Cs-Sb-O$_2$ layer achieves significant enhancement in lifetime near the photoemission threshold.

\setlength{\tabcolsep}{4pt}	
\begin{table*}
	\centering
	
	\begin{tabular}{l|llll}
		\hline \hline \\
		\multirow{2}{5em}{Activation method} & \multirow{2}{5em}{Initial QE at 780 nm}            & \multirow{2}{5em}{Final QE at 780 nm}               & 
		\multirow{2}{7em}{Lifetime estimate (hour)} & \multirow{2}{6em}{Improvement factor} \\
		 \\
		\hline
		\\Cs-O$_2$ & $3.3 \times 10^{-2}$              &    $1.4 \times 10^{-3}$           & 
		$15$    &$\times 1$                                                                \\
		Cs-O$_2$ + Cs-Sb			 & $9.5 \times 10^{-3}$ & $2.9 \times 10^{-3}$	& 
		$41$ &$\times 2.7$\\
		Cs-Sb + Cs-O$_2$    & $3.8 \times 10^{-3}$              & $1.1 \times 10^{-3}$      & 
		$70$        & $\times 4.6$                                                                   \\
	
		Cs-Sb-O$_2$				& $1.0 \times 10^{-2}$			& $7.0 \times 10^{-3}$	&
		$104 $	&$ \times 6.8$	\\

		\hline \hline                                                                             
	\end{tabular}
	\caption{Lifetimes at 780 nm estimated by comparing initial QE and final QE after the QE degradation measurement in Fig.~\ref{fig_lifetime}.}
	\label{table2}
\end{table*}


\begin{figure}
	\includegraphics*[width=230pt]{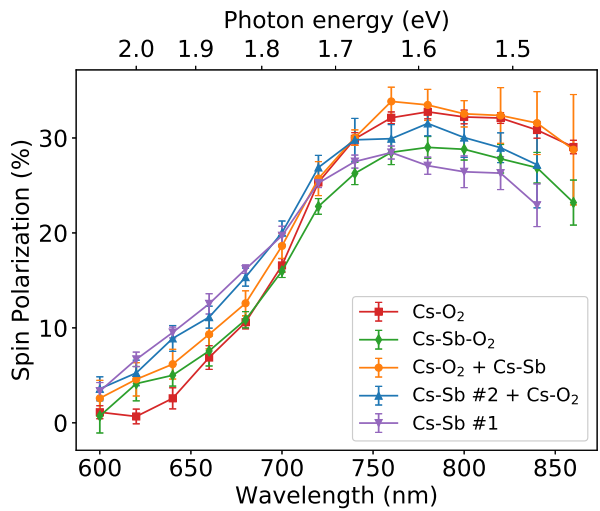}
	\caption{Spin polarization of photoemitted electron from GaAs activated by Cs, Sb, and O$_2$.}
	\label{fig_polarization}
\end{figure}

Spin polarization of photoemitted electrons was measured by the Mott polarimeter described in Refs.~\onlinecite{Mulhollan1994,gaas_cste}.
Monochromatic light was used to produce circularly polarized light directed at normal incidence to the sample surface.
Longitudinally spin polarized electrons were bent $90^\circ$ by electrostatic lenses and Mott scattered with 20 keV at a tungsten target which was calibrated to have 0.18 Sherman function.\cite{Mulhollan1994,gaas_cste} 
Photoelectron spin polarization was measured as a function of wavelength for each sample in Fig.~\ref{fig_polarization}. The maximum spin polarization varied up to 5\% which corresponds to about 3 standard deviations.
The detailed study\cite{LucaPaper} on the samples activated with the Cs-Sb-O$_2$ method reveals a significant spin depolarization as thickness of the Sb layer grows.
 
 

\begin{figure}
	\includegraphics*[width=230pt]{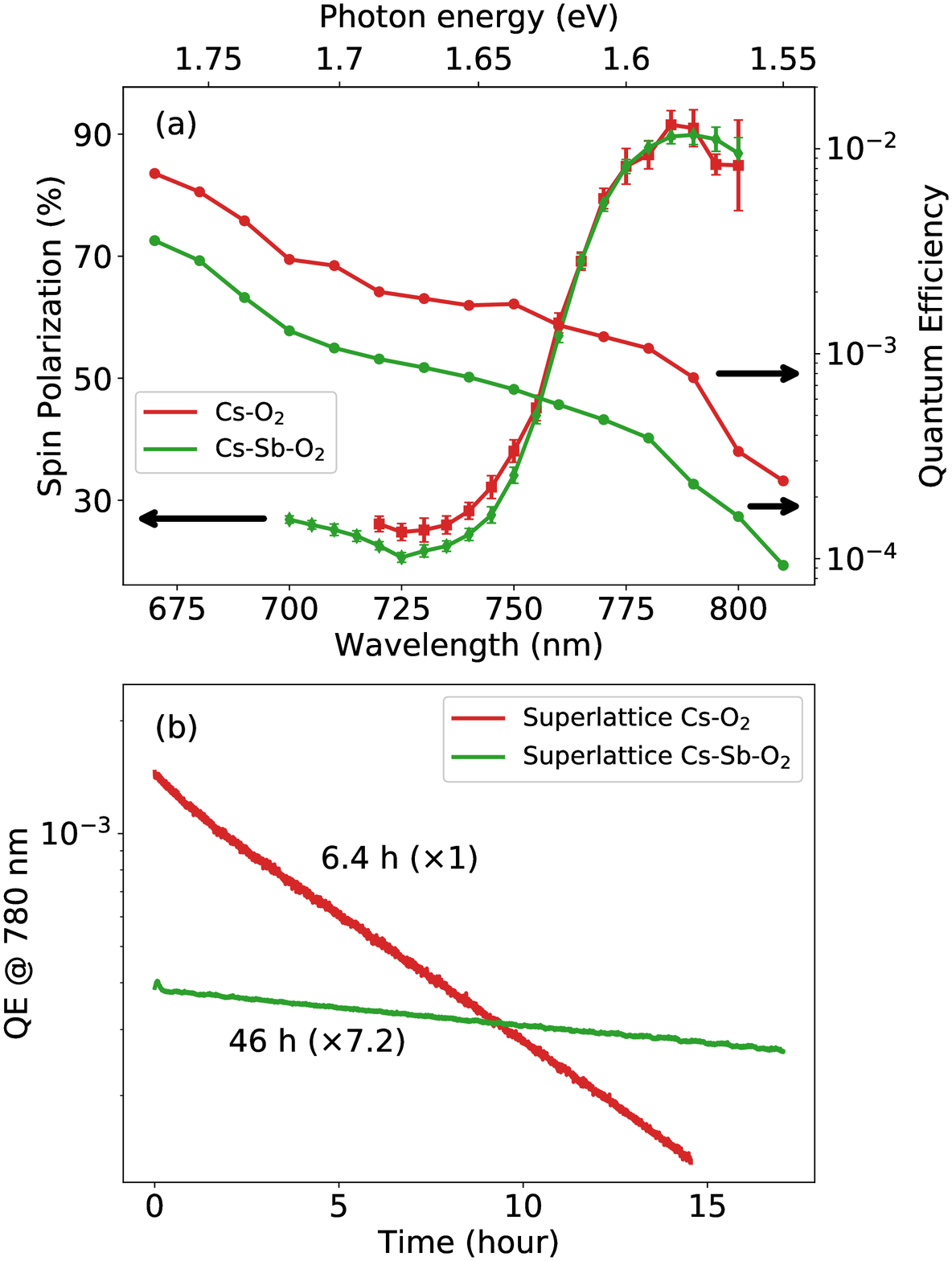}
	\caption{Spin polarization, quantum efficiency, and lifetime measurement of GaAs/GaAsP superlattice sample activated by codeposition of Cs-Sb-O$_2$ and standard Cs-O$_2$.}
	\label{fig_superlattice}
	
\end{figure}

Since the Cs-Sb-O$_2$ codeposited sample showed both the highest QE at the photoemission threshold among Sb deposited samples and  a significantly improved lifetime, we chose the codeposition method to activate a GaAs/GaAsP superlattice sample\cite{Maruyama2004} (Fig.~\ref{fig_superlattice}). Compared to the bulk samples, a half the amount of Sb is deposited (1.25 \AA) to minimize depolarization at the activation layer.\cite{LucaPaper} Both samples showed $\sim$ 90 \% maximum spin polarization with agreement within one standard deviation. The QE at 780 nm was a factor of 3 smaller for the semiconductor activated sample. QE was continuously measured in Fig.~\ref{fig_superlattice} (b) for lifetime estimation at 780 nm, the wavelength commonly used for a highly spin polarized photoemission.
The lifetime showed a factor of 7 improvement, which is similar to the bulk results in Table.~\ref{table2} that were activated with twice the amount of Sb.

\section{Discussion}

The sample prepared with method (c) (where the Cs-O$_2$ is deposited on the surface of GaAs under the Cs-Sb layer) has a higher QE than the sample prepared with method (b), where the Cs-O$_2$ is deposited above the Cs-Sb layer. This suggests that despite the small thickness of the activating layers, segregation can take place such that the intermediate Cs-O$_2$ layer retains the strong electric dipole enhancing the transmission from GaAs into the Cs-Sb layer.
Additionally, the QE near threshold of the sample exposed to Cs-O$_2$ after depositing Cs-Sb is larger than the one activated only with Cs-Sb.
Studies on the nature of the GaAs/Cs-O activation layer indicate that the dipole layer consists of a Cs$^+$-O$^{2-}$-Cs$^+$ sandwich on a O-GaAs layer, and the strong double dipole layer results in a higher QE compared to Cs only activation.\cite{Su1983} On the other hand, studies on the oxidation of alkali antimonide photocathodes show that, during oxidation, the films segregate in a alkali-oxide rich surface layer on top of the alkali antimonide layer. \cite{Soriano1993} In particular, at low oxygen exposures (less than 10 L for Cs$_{2.5}$K$_{0.5}$Sb), \cite{Soriano1993} the top surface is mainly composed of a Cs suboxide (Cs$_{11}$O$_3$) \cite{GaldiNAPAC} that is known to promote photoemission, in particular at long wavelengths.\cite{Bates1981}

The above discussion suggests that the sample activated with Cs-Sb-O$_2$ should indeed have the the largest QE among the alternative methods. In this case, the continuous oxygen exposure can result in the formation of strong electric dipole at both the interface with GaAs and at the surface of Cs-Sb that can favor both electron tunneling from GaAs to Cs-Sb and electron emission from Cs-Sb to vacuum. The oxygen dosed on the samples during codeposition can be estimated to be of the order of 50 L - 75 L; since Cs is provided during the oxygen exposure and after the closure of the oxygen leak valve, it is reasonable to expect the formation of Cs-rich suboxides on the sample surface.



\begin{figure}
	\includegraphics*[width=230pt]{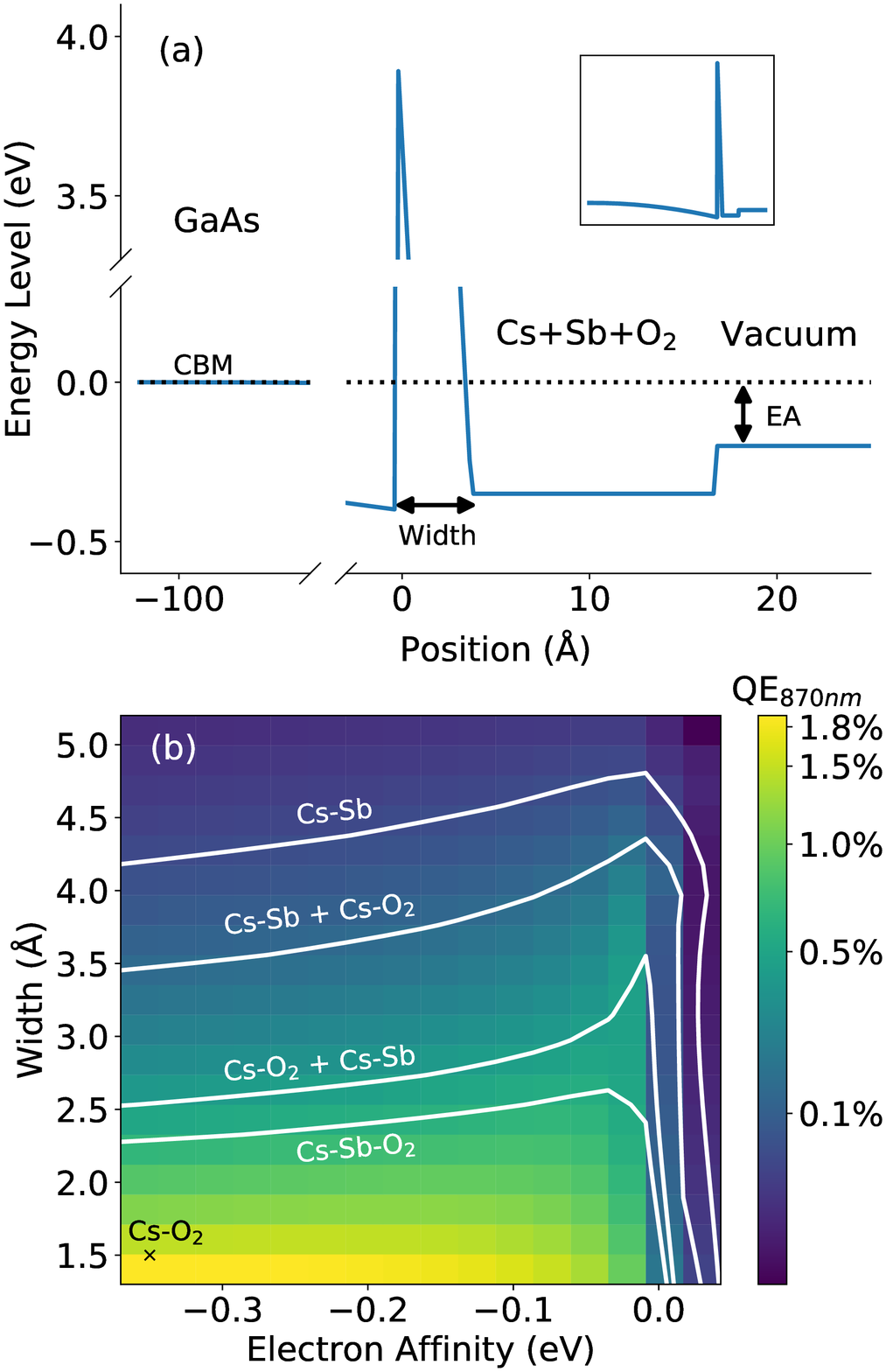}
	\caption{Numerical simulation of quantum efficiency under various potential barrier shapes. (a) Potential barriers were constructed for 17\AA\ thick semiconductor activation layer to calculate transmission probabilities. (b) Width and EA are varied to compute threshold QE. The black mark corresponds to the standard Cs-O$_2$ activation sample. Potential barrier parameters in Ref.~\onlinecite{Vergara1999} were used. White contour lines are sets of parameters that reproduce experimentally measured QEs.}
	\label{fig_potentialBarrier}
	
\end{figure}

Photoexcited electrons experience a potential barrier at the GaAs surface in the process of emission into the vacuum. Although various models of the potential barrier were proposed, fundamental understanding of the nature of the barrier is lacking.\cite{Karkare2013, Jin2014} Rectangular,\cite{Fisher1972} triangular,\cite{Vergara1999} quadratic,\cite{Subashiev2000} and double triangular\cite{Jin2014} shapes have been proposed in previous studies to reproduce various experimental observations. In Fig.~\ref{fig_potentialBarrier}, a simple potential barrier was constructed to numerically investigate how different methods of oxygen exposure during Cs-Sb growths can affect the shape of the potential barrier and QE as a result.

In the bulk of GaAs, we implement quadratic band-bending near the surface with a 100 \AA\ band bending region and 0.4 eV band bending magnitude.\cite{Karkare2013} A simple triangular potential barrier was placed at the interface between GaAs and the activation layer. The height of the triangle was fixed at 4 eV above the conduction band minimum (CBM) of GaAs\cite{Vergara1999} while the width was varied throughout the simulations. The variation of the width can be interpreted as a change in dipole moment strength at the interface.
The bottom of the potential well between the triangular barrier and the vacuum interface was fixed at 0.35 eV below the conduction band minimum.\cite{Uebbing1970, Jin2014, Bakin2015}
The vacuum interface is positioned at 17 \AA\ from the GaAs surface which is the estimated thickness of the activation layer.

The transmission probabilities ($P$) of constructed potential barriers are calculated by the propagation matrix approach\cite{Karkare2013,levi2006applied,gilmore2004elementary} and used to compute QE numerically with the following equation:\cite{3step}
\begin{equation}
	\text{QE} = (1-R) F\, \frac{\int^\infty_0 dE\, f(E) \int^1_0 d(\cos\theta) P(E \cos^2 \theta)}{\int^\infty_0 dE\, f(E) \int^1_{-1} d(\cos\theta)}.
\end{equation}
Here, $R$ is the optical reflectance of GaAs, $F$ is a scattering contribution (from holes, impurities, and phonons)\cite{Karkare2013} which can be approximated to be independent of energy near the photoemission threshold,\cite{3step} $\theta$ is the angle between the electron velocity and the surface normal, and $f(E)$ is a Fermi-Dirac distribution of photoexcited electrons:
\begin{equation}
	f(E) = \frac{1}{1+e^{[E-(h\nu -E_g)]/kT}}
\end{equation}
where $h\nu$ is the photon energy, $E_g$ is the GaAs band gap, $k$ is the Boltzmann constant, and $T$ is the room temperature. Additional scatterings at the semiconductor activation layer can be considered mostly elastic electron-electron collisions, therefore it is ignored.\cite{LucaPaper}

In Fig.~\ref{fig_potentialBarrier} (b), QEs are calculated for various sets of the triangle width and electron affinity. The photoemission threshold wavelength 870 nm was used, and the scattering term $F$ was set to be 0.21 to match the experimentally measured value of Cs-O$_2$ activated sample to the calculated one. The width of triangular potential barrier of Cs-O$_2$ activation layer is reported to be 1.5 \AA\cite{Vergara1999} with 0.35 eV NEA.\cite{Jin2014, Bakin2015,Uebbing1970} The white contour line for each growth method represent the set of parameters that reproduce the experimentally measured QE at 870 nm.
The contour lines suggest when NEA is greater than 0.1 eV, the efficiency is roughly independent of NEA while the barrier width plays critical role on QE. Considering that a small width represents strong dipole moment on the surface, activation with oxygen at the interface between bulk GaAs and activation layer is essential to achieve a high QE.
Thus, this simple model can explain why samples activated with oxygen before and during the Cs-Sb growth showed greater improvement in QE compared to the sample exposed to oxygen after Sb deposition which is related to greater NEA at the vacuum interface.


\section{Conclusion}

We have investigated the NEA activation of GaAs samples with Cs, Sb and O$_2$ in various recipes. Of the recipes attempted, we found that the codeposition method can achieve the highest QE ($\sim 1\%$ at 780 nm) with a factor of 10 and 7 improvement in lifetime at 505 nm and 780 nm, respectively, compared to the conventional Cs-O$_2$ activation.
Similar results were obtained for the high polarization GaAs/GaAsP superlattice sample. This sample showed a factor of 7 improvement in lifetime at 780 nm without any depolarization at a cost of a factor of 3 smaller QE.
A simple numerical model was proposed to estimate QE for various shapes of the potential barrier at the relatively thick semiconductor activation layer. According to the model, the dipole moment strength at the GaAs interface with the activation layer is critical to achieve a high QE at the photoemission threshold. This explains the significant improvement in QE we observed when the samples were exposed to oxygen before activation layer growths.

Future work will involve surface characterization of the activation layer to understand the chemical and structural composition. This information can be used to properly model the heterojunction band structures with density functional theory. Monte Carlo techniques can be considered to model the electron spin transport and to study the depolarization mechanisms. Furthermore (and perhaps most critically), we plan to test the performance of these unconventional activation layers in a real high voltage, high current DC gun environment in the future.

\section{Acknowledgments}
This work was supported by the Department of Energy under Grant No. DE-SC0019122 and National Science Foundation under Grant No. PHY-1549132.
Authors would like to acknowledge Marcy Stutzman, Matt Poelker, Joe Grames, and William DeBenedetti for valuable discussions, help in preparing GaAs samples, and providing GaAs/GaAsP superlattice samples.


\bibliographystyle{unsrt}
\bibliography{reference} 

\end{document}